\documentclass{sigchi}

\usepackage{balance}  % to better equalize the last page
\usepackage{graphics} % for EPS, load graphicx instead 
\usepackage[T1]{fontenc}
\usepackage{algpseudocode}
\usepackage{txfonts}
\usepackage{mathptmx}
\usepackage[pdftex]{hyperref}
\usepackage{color}
\usepackage{booktabs}
\usepackage{textcomp}
\usepackage{microtype} % Improved Tracking and Kerning
\usepackage{ccicons}  % Cite your images correctly!
\usepackage{todonotes}

\def\plaintitle{InfraNotes: Inconspicuous Handwritten Trajectory Tracking for Lecture Note Recording with Infrared Sensors}

\def\emptyauthor{}
\def\plainkeywords{Lecture note, Handwritten, LeapMotion, OCR, Infrared, Inconspicuous}

\makeatletter
\def\url@leostyle{%
  \@ifundefined{selectfont}{
    \def\UrlFont{\sf}
  }{
    \def\UrlFont{\small\bf\ttfamily}
  }}
\makeatother
\def\pprw{8.5in}
\def\pprh{11in}

\definecolor{linkColor}{RGB}{6,125,233}
\hypersetup{%
  pdftitle={\plaintitle},
  pdfauthor={\emptyauthor},
  pdfkeywords={\plainkeywords},
  bookmarksnumbered,
  pdfstartview={FitH},
  colorlinks,
  citecolor=black,
  filecolor=black,
  linkcolor=black,
  urlcolor=linkColor,
  breaklinks=true,
}

\begin{document}

\title{\plaintitle}

\numberofauthors{1}
\author{%
  \alignauthor{Steve Chang\\
    \affaddr{University of California, Los Angeles}\\
    \email{USA}}\\
}

\maketitle

\begin{abstract}
  Lecture notes are important for students to review and understand the key points in the class. Unfortunately, the students often miss or lose part of the lecture notes. In this paper, we design and implement an infrared sensor based system, InfraNotes, to automatically record the notes on the board by sensing and analyzing hand gestures of the lecturer. Compared with existing techniques, our system does not require special accessories with lecturers such as sensor-facilitated pens, writing surfaces or the video-taping infrastructure. Instead, it only has an infrared-sensor module on the eraser holder of black/white board to capture handwritten trajectories. With a lightweight framework for handwritten trajectory processing, clear lecture notes can be generated automatically. We evaluate the quality of lecture notes by three standard character recognition techniques. The results indicate that InfraNotes is a promising solution to create clear and complete lectures to promote the education.
\end{abstract}

\category{H.5.m.}{Information Interfaces and Presentation
  (e.g. HCI)}{Miscellaneous}{}{}

\keywords{\plainkeywords}

\section{Introduction}

Learning from class notes is an important process for students. Taking lecture notes is an active learning method to keep students concentrating during lecture and help the review process. However, students commonly miss parts of class notes due to insufficient note taking speed, being late to class, or the occasional interruption. In fact, incomplete class notes is one of the reasons that preventing students from efficiently reviewing the knowledge after class. As a result, more and more students record the lectures using cameras and replay them later for reviewing. On the other hand, some schools require instructors to post their lecture notes online before the class. Nevertheless, if lecturers make any impromptu updates in their lecture notes or answers unexpected questions on the black/white board, these modifications will not be included in the prior-posted notes. Therefore, more researchers tend to invent technologies to help on automatic lecture notes recording.

\begin{figure}
\centering
\includegraphics[width=0.20\textwidth]{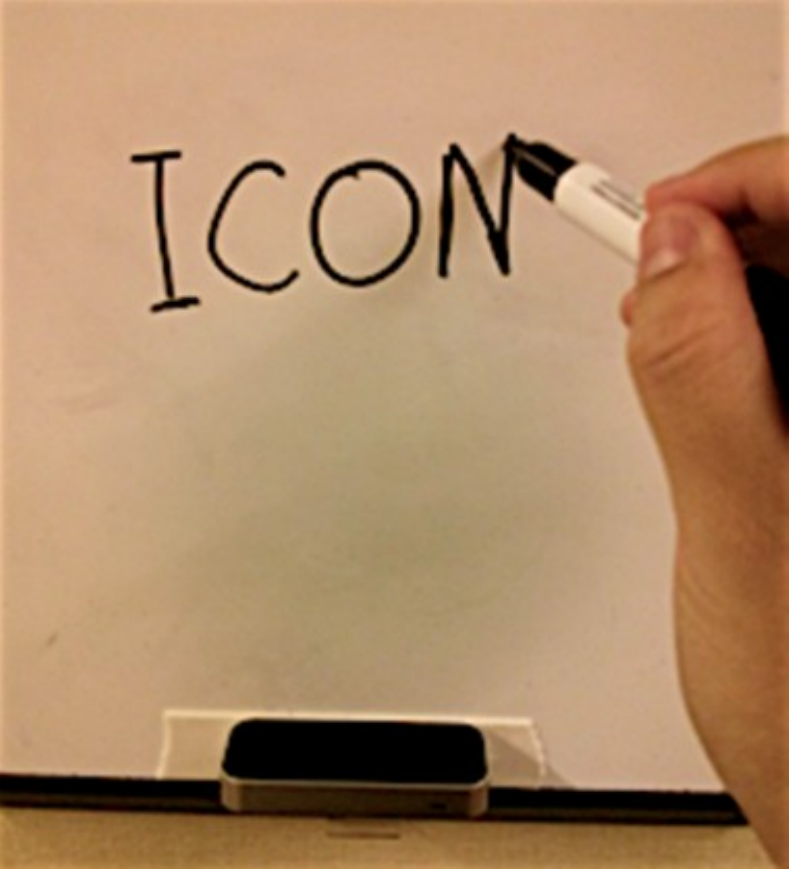}
\includegraphics[width=0.20\textwidth]{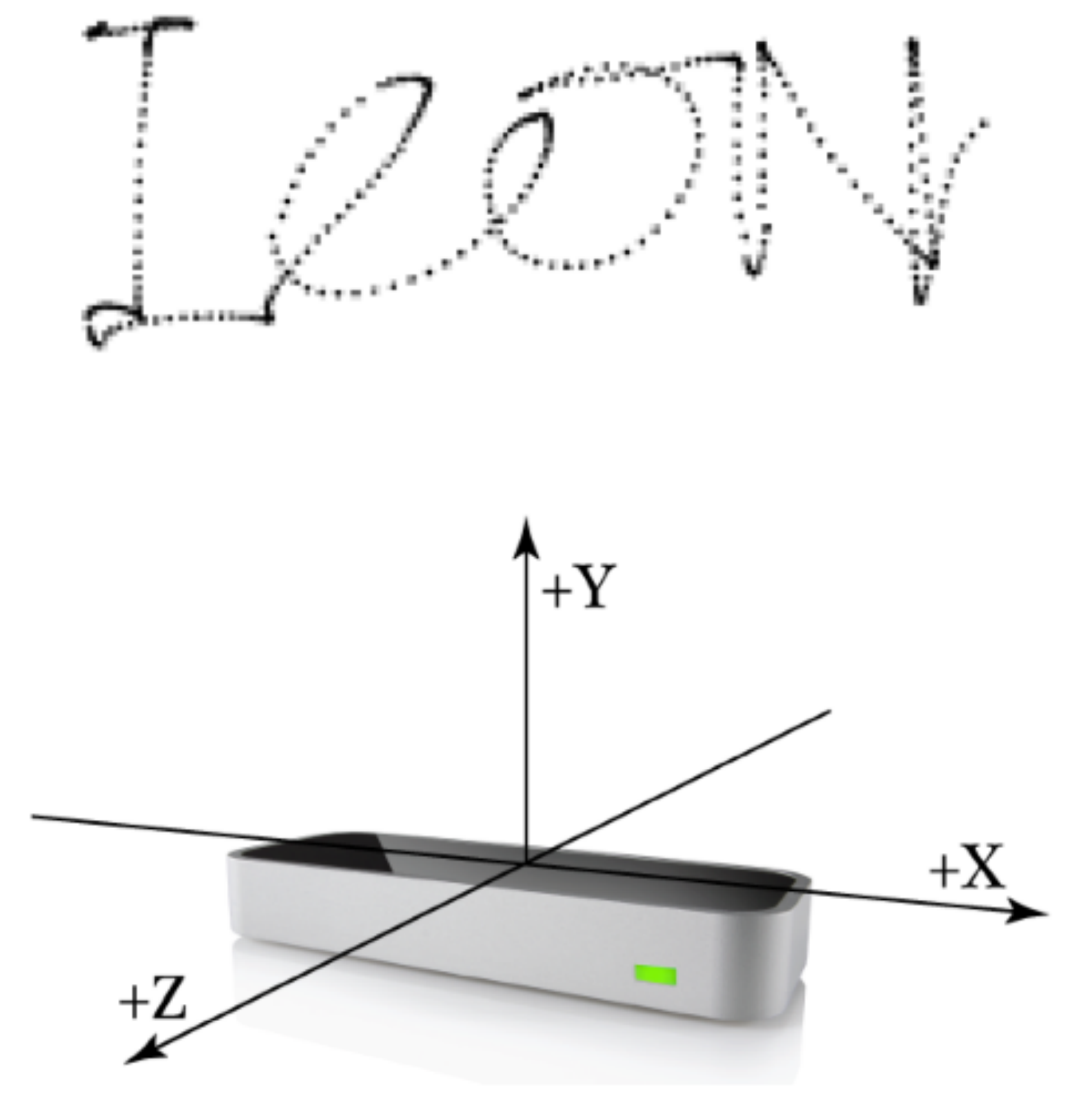}
\label{fig.Figure1}
\caption{The infrared sensing device setup is shown in left figure (Y-axis is the direction of infrared sensors). A comparison of written characters on the board and the handwritten trajectories from infrared sensors are plotted in right figure.}
\end{figure}

Two straightforward solutions are proposed: the first is to extract notes from videos, and the second is to use sensors (touch surface or special designed pens) to record writing movement, commonly called pen and paper computing. Our system design belongs to the second one. Infrared sensors (Fig. 1) are used to capture handwritten trajectories. This design is based on tracking the geometric information, such as the distance and location between a writing tool and a black/white board, and writing movements can be recorded with upward projecting infrared sensors. In comparison, a video recording system is usually expensive and requires spacious class room space, while using InfraNotes to record writing trajectories is more affordable and convenient to set up. Furthermore, there is no special calibration needed to ensure no line-of-sight blocking and incomplete field of view issues. It can also avoid to handle challenging technical problem of extracting non-duplicated class notes from a sequence of videos. Therefore, several sensor-based note taking systems are introduced by tracking handwritten movements with a touch surface or a sensor-equipped pen. Touch surface is also a popular system. Lecturers can use their fingers or a touch pen to write material on a touch surface. All written information can be automatically recorded and saved as downloadable files. Still, some setup is necessary, replacing black/white board with a touch surface is troublesome, and traditional chalk or markers cannot be used on such touch surfaces. Using a sensors equipped pen can be a lightweight solution, but still lecturers are forced to adopt those special designed pens. We are specially interested to explore the possibility of combining the benefit of video taping system (less interference with lecturers in teaching) and the benefit of pen-based sensors equipped devices (no special deployment and setup are required) for inconspicuously supporting common lecture activities; for example, lecturers may write, erase, overwrite existing notes and student may want to search notes or even search taped lecture clips by keywords learned from a lecture for reviewing course materials. We design a novel handwritten tracking system with special data processing methods based on observed English writing behaviors and propose a feasible solution to support all mentioned common lecture activities for both lecturers and students.

Here we would like to highlight the novelties and contribution in this work:
\begin{enumerate}
\item Designing a novel infrared-sensor based system to capture the handwritten trajectory;
\item Developing a complete processing framework from data processing, notes generation, modification, assembling, to bidirectional search;
\item Evaluating notes quality/readability by standard character recognition techniques, including OCR, Stroke Heuristics, and Stroke Tables.
\end{enumerate}

\section{Related Work}

Before we proceed to system design, some prerequisite terminologies are explained in this section: hand-writing recognition and existing sensor assisted notes recorders. Hand-writing recognition techniques are utilized to process sensor data stream and to evaluate the readability of the notes content.

\subsection{Overview Sensor Assisted Notes Recorders}

Some universities recorded lecture videos, therefore, a variety of researches focus on photo/video analysis in order to generate class notes. One practical solution was blackboard segmentation \cite{Cite01}. This system first analyzed video to segment written regions on blackboard. Then it exported these regions of notes in photo format. This solution was effective especially when processing lectures in the past. However, it could only generate photo, which was not a search-friendly content. Another method \cite{Cite02,Cite03} was to use a camera scanning whole white board with several photos. Then it stitched those images together as a panorama and eliminate background and keep enhanced notes image. This solution was practical and generated readable notes image. There was also a system \cite{Cite04} which could recognize handwriting text from these photo by OCR. However, these two solutions could not automatically generate notes with time line in every line. In real settings, this system might miss content if it missed a single photo. The modification on notes also could not be preserved. Most significantly, the writer must not block the notes on the board, which was hard to achieve during lecture.

Due to drawbacks of photo/video analysis, companies released interactive whiteboard with huge touch screen. A commercial product, IR Touch \cite{Cite05}, is a 65 inch LCD screen with infrared frame to detect touch events. Touch screens can provide fast response, and nowadays users get used to touch screen from Apple products. However, a screen can only achieve a relatively small range from 60 to 100 inches. Most importantly, the large touch screens are expensive, currently priced \$3000-5000 based on screen sizes.

Compete with high price of touch screen, current researchers have investigated methods to automatically record notes using cheaper sensors. Microsoft Kinect RGB-D camera \cite{Cite06} could provide depth information based on its infrared camera. It stimulated researchers to innovate in remote control\cite{Cite07} and education\cite{Cite08}. The Low-Cost Efficient Interactive Whiteboard \cite{Cite09} used a Kinect to detect and track user hand movement on the whiteboard. Unfortunately, Kinect cannot provide accurate enough depth resolution when the hand is near the board, so this work combined visual data with depth information to track hand movements. Another revolutionary sensor, Nintendo Wiimote \cite{Cite10} could provide relative position from a fixed IR emitter. Researchers explored its potential in 3D User Interface\cite{Cite11}, Medical Data Visualization\cite{Cite12} and Human Computer Interaction\cite{Cite13}. Lee et al. designed an interactive projection solution \cite{Cite14}. In this system, a writing tool was equipped with an IR emitter and its position was tracked by using a fixed Wiimote on its field of view. A computer screen was projected onto a wall, the IR emitter writting tool was tracked accordingly and it allowed the user to write onto the projected computer screen image. Unfortunately, the Wiimote's field of view could be blocked by user body in daily usage, and therefore the system loses valualbe tracking information.

In order to fix blocking problem, researchers and companies came up with solutions with special writing tools. ThumbsUp \cite{huang2015leveraging} recognizes writing gestures by EMG sensors, which will not be visually blocked. Ee-Class \cite{Cite15} was an interactive education support system. Students could write a letter with Wiimote in the air to answer questions. However, this system was not designed to write continuously. Also, the students had to perform special actions (pause or press button) to segment strokes. Although it could not achieve full functionality, writing a letter was a good start of classnotes taker. Researchers also realized the potential of sensors inside cellphones. PhonePoint Pen \cite{Cite16} used the built-in accelerometer to track phone movement in the air, and then recognized human writing by reconstructing the relative position from acceleration with noise. This system used threshold to segment strokes, and then used grammar tree to recognize letters. However, its grammar tree forced writing order to reduce ambiguity in letters with the same strokes. Also it required users to hold the phone in a fixed pose, in order to fix patterns in accelerometer reading. Moreover, its average writing speed, 3s/letter, was not acceptable to write long content like lecture notes. According to a comprehensive survey to pen-and-paper user interface \cite{Cite17}, the most suitable solution to notes taking was ActivBoard \cite{Cite18}. This product did not use infrared touch screen. Instead, it used special ActivPen to interact with whiteboard. The sensor inside the pen could track the relative position of writing on the board. Unfortunately, the writer needed to get used to ActivPen, which was thicker and heavier than marker or chalk. Moreover, the need of battery in ActivPen can be troublesome in actual working environment. Proper Running Posture Guide \cite{zhang2013proper} showed the potential of on body sensor to recognize activities; since it does not rely on visual signal, the actions will not be blocked by the user's body.

\subsection{Overview Hand-writing Recognition Research}

Hand writing recognition is the task of transforming a language represented in its spatial form of graphical marks into its symbolic representation. There are two kinds of handwriting input, on-line and off-line \cite{Cite19}. On-line handwriting input maintains the time series of writing points, order of strokes and additional information about pen tip (velocity, acceleration). For example, handwriting input methods on cell phones and tablets receive on-line handwriting input when users touch the screen. Preprocessing of on-line recognition includes noise removal, stroke and character segmentation. Off-line handwriting input only preserves images of the completed on-board writing area. For example, banks recognize handwritten amounts on checks. Preprocessing of off-line recognition includes setting thresholds to extract writing points, removal of noise, segmentation of writing lines, and finally segmentation of characters and words. Our system is a type of on-line handwriting input system and records time stamps of each points on the handwritten trajectory. However, the sensor tracks both on-board and off-board movement of writing tools. Thus, we need classification after segmentation to determine which strokes are belong to on-board writing or off-board hand movement.

\section{System Overview}

\begin{figure}
\centering
\includegraphics[width=0.48\textwidth]{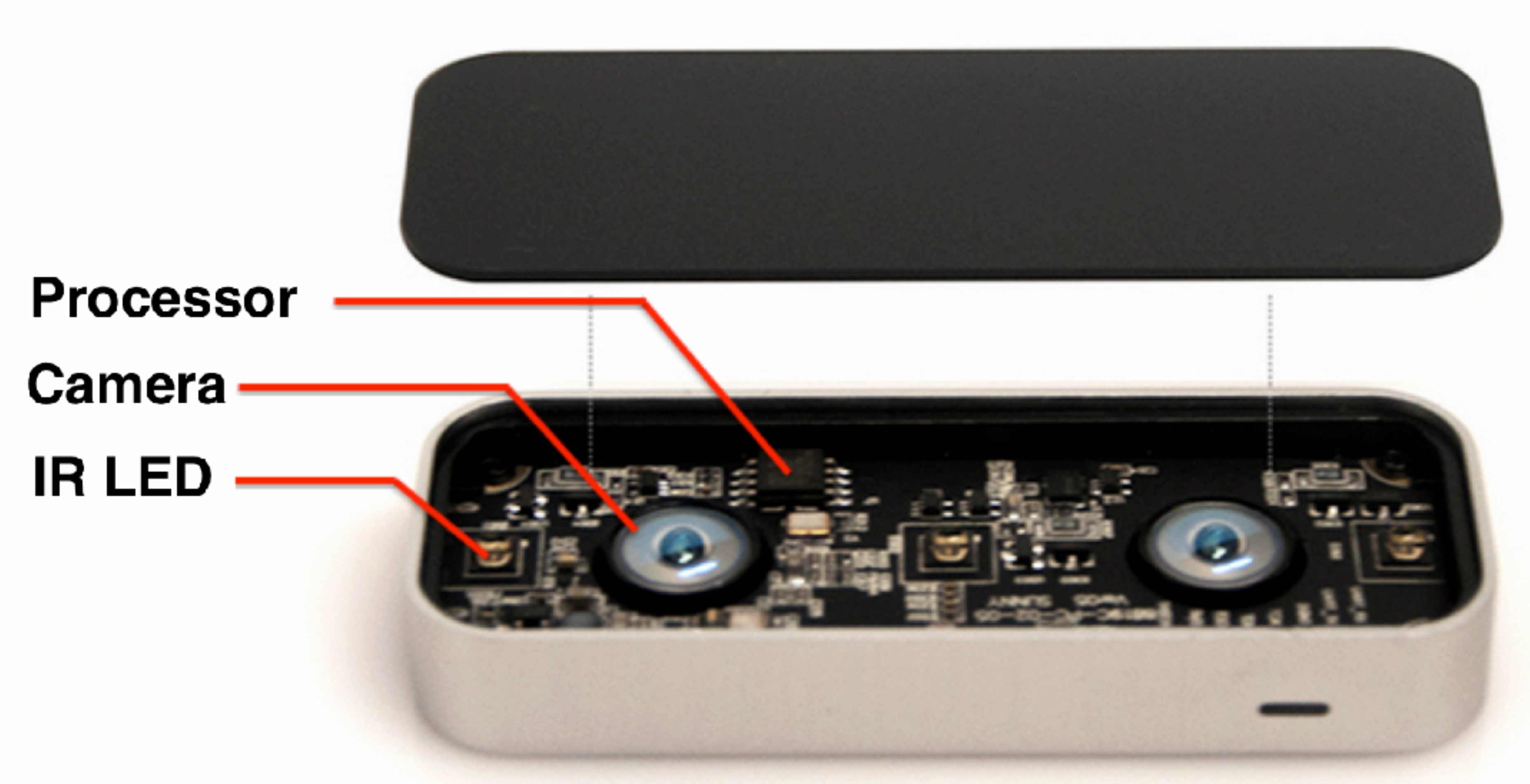}
\caption{LeapMotion: Infrared Sensor System}
\label{fig.Figure2}
\end{figure}

Hardware characteristics and deployment techniques and software flowchart are described in this section. Through literature reviews, we understand the pros and cons of the sensor-equipped devices. By taking advantage of classroom environment, deploying our proposed system can be simple and efficient. A sequence of software processing steps further automate the procedures of creating lecture notes without interventions of lecturers and students.

\subsection{LeapMotion: Infrared Sensor System}

\begin{figure*}[htb]
\centering
\includegraphics[width=1.0\textwidth]{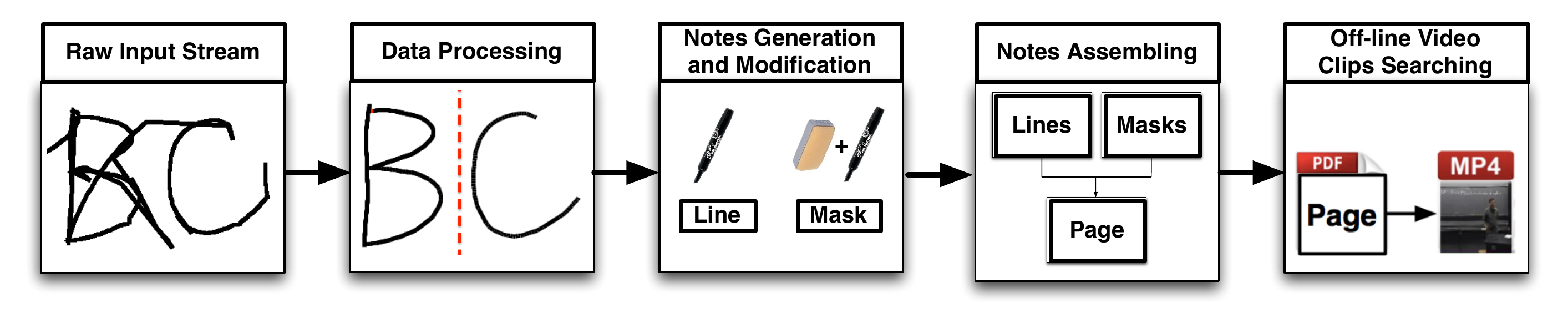}
\caption{System Flowchart}
\label{fig.Figure3}
\end{figure*}

LeapMotion\cite{Cite20} is a brand new dual camera tracking system that uses infrared light to detect location, velocity and orientation information of targeted objects. It equips with infrared leds, infrared sensors, and an embedded processor (Fig. 2). By receiving reflected infrared signals from the objects with emitting range, it can compute the \mbox{3-D} position of the objects relative to the device. It is designed to precisely track movements of thin objects such as fingers or pens. Moreover, multiple objects (fingers) can be tracked at the same time. Its sampling rate can range from 30 to 200 frames per sec based on CPU performance, and it will normally cost less than 2\% CPU power. Therefore, developers have mainly focused on using this sensor to track fingers movements and build gesture control interfaces of computers. Nevertheless, we realize its potential in assisting lecture notes creation in classroom environment. A flat eraser-holder commonly accompanies with black/white boards in a classroom. Chalks, markers, and erasers are placed on top of the holder. If we place the sensor on the holder and project upward, when a lecturer writes notes, his/her writing movement must be observed. This observation inspires us to use upward projected LeapMotion sensor for class note collection. This affordable (\$70) sensor has theoretical 0.01mm accuracy in \mbox{3-D} position, which is 200 times more accurate than anything else on the market, at any price point"\cite{Cite20} such as Kinect. Ideally, all movements enter the field of view of the sensor can be easily captured and recorded under the claimed precision. However, in real deployment, we tested its accuracy by steadily placing a marker tip on the board for 10 seconds. We verified the precision in millimeters with rulers, but we found that it has 0.35mm variance in xy plane and 0.14mm in z depth. This accuracy has negligible effect on general gesture control of computer control, as gesture sequences are much more important than the accurate position. However, this accuracy can highly affect on handwritten movement collection, because identifying whether the collected movement is writing in the air or writing on the board becomes non-trivial. The variance of z direction will dominate the "writing or not" judgements and we do not want to require lecturers to move chalks or pens from board for a large enough distance when they are not writing. In addition, in system design section, we will see some examples which show that there are several transition strokes (the movement between two on-board strokes) when a character is written. These transition strokes highly affect the readability of the collected handwritten trajectories.

\subsection{InfraNotes System Flowchart}

Fig 3. presents the InfraNotes system flowchart. There are four main components in the software architecture: Data Processing, Notes Generation/Modification, Notes assembling, and Bidirectional (lecture notes and lecture clips) search. Raw data stream returned from LeapMotion contains position, velocity, and orientation information of every points within the handwritten trajectories. Segmentation, classification, and grouping algorithms are applied on the raw data based on this collected information which implies human English writing behaviors and conventions. All alphabets written in a row are stored as an image after algorithmically cleaning. Two image types (Line and Mask)are assigned, depending on whether the lecturer writes on a new row or simply modify contents in an existing row. Mask is a patch of the existing Line. The content in Mask will overlay on top of an existing Line when a series of Lines and Masks are assembled to form pages of lecture notes. Each page of lecture notes is created when the lecturer starts a new column and these pages can be used to index video clips with stored time stamps of each Line, Mask, and Page. Point series to characters conversion can be done with OCR and alphabet stroke table and it can be used to support keyword-based notes or clips search across all text identifiable notes.

\section{System Design}

This section illuminated the five main modules of the InfraNotes system design: data processing, class note generation, class note updating, class note assembling, and off-line video segments searching.

\subsection{Data Processing}

\begin{figure}[h!]
\centering
\includegraphics[width=0.48\textwidth]{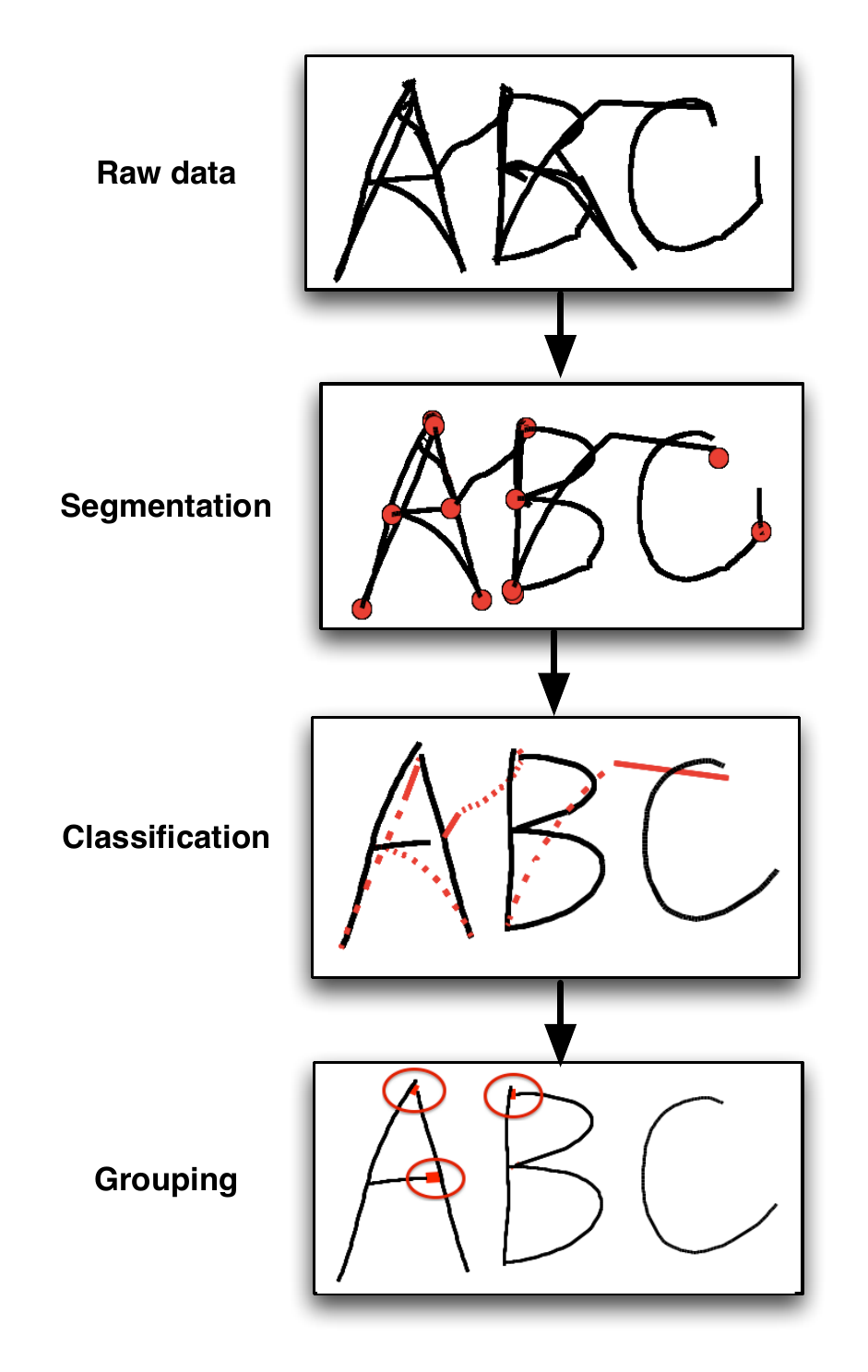}
\caption{Data processing algorithm flow is split into three parts: segmentation, classification and grouping. Segmentation step detects the start and end of each stroke from the raw data stream. Classification step determines whether a given stroke is an on-board handwritten trajectory. Grouping step groups and connects on-board stokes into an alphabet}
\label{fig.Figure4}
\end{figure}

Raw data stream is composed by a sequence of points with writing tool's position, velocity, orientation and timestamp. A subset of points can form a stroke, then those strokes form a character, then a word, and then a sentence. Data processing steps shown in Fig 4 aim to segment data stream into strokes, remove the off-board movement, and group the on-board strokes into a character. The heuristics used for segment data points are oriented from our observations of English handwritten behaviors and conventions. However, the classification is not trivial. LeapMotion sensor loyally returns all points, including both on-board writing and off-board movement of writing tool. Due to the accuracy limitation of LeapMotion, it is improper to classify on-board and off-board trajectories simply based on z-axis coordinates. Since single point information is not sufficient to make appropriate classification, we attempt to segment a set of points into a stroke, and make on-board or off-board classification for each stroke. After segmentation and classification, on-board handwritten strokes are ready for grouping. Grouping step is optional when producing image-based lecture notes, but it is useful for notes quality evaluation and character recognition in the keyword searching service.

\subsubsection{Segmentation}

The goal of segmentation is to identify individual strokes from the raw data stream based on human English handwritten behaviors and conventions. Three types of features can identify the start and end points of a stroke, including slow start and slow end in XY-plane, big jump in Z-direction speed diagram, and sharp angle transition in a handwritten sequence.

\begin{enumerate}
\item Slow start and slow end in XY-plane:
When a writer attempts to write a stroke, he/she will start a stroke from slow speed, then speed up to a constant speed, and then slow down when the end of a stroke is reached.
\item Big jump in a Z-direction speed diagram:
If the following stroke is not connected with the current written stroke, a writer should lift the writing tool so that he/she can move to the start of the next stroke. Thus, the speed impulse (big jump) in Z-direction can be used to segment between two consecutive strokes.
\item Sharp angle transition in a handwritten sequence:
If a handwriting velocity sequences contains a sharp direction transition, the hand-writing sequence should be split into basic smooth strokes to facilitate character recognition.
\end{enumerate}

Hand-writing movement of a character "A" can be split into five strokes from s0 to s4 marked in Fig. 5. Raw data returned by LeapMotion loyally record all handwritten movements, including s1 and s3 which are the strokes written in the air. Nevertheless, we can clearly identify each stroke by picking low XY-plane speed points as the start or end points. When writing a stroke, the XY-plane speed is much larger than 50mm/s, and the  low writing speed distinguishs two strokes, as we can see in Fig 5.

\begin{figure}[h!]
\centering
\includegraphics[width=0.48\textwidth]{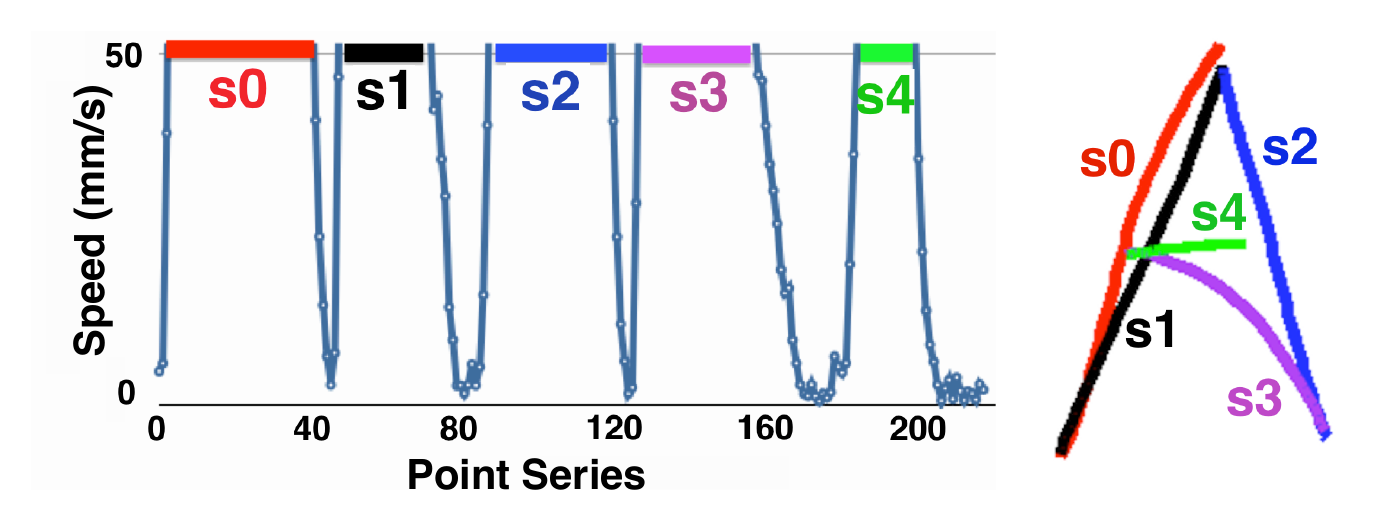}
\caption{Character "A" segmentation is based on XY-plane Speed. Segmented strokes are marked with different colors which matches with the XY-plane speed diagram. Every low speed point in the diagram can be a start and end point of a stroke.}
\label{fig.Figure5}
\end{figure}

Fig. 6 shows a Z-direction speed diagram when a writer needs to move to next start point of a stroke by lifting his/her writing tool. A big jump can be observed in the s1 period. Once s1 is identified, its neighbor strokes s0 and s2 are treated as separate strokes. In this example, s0 and s2 are even belong to two different words, "K" and "L". This feature works on writing two strokes which are not connected to each other within an alphabet as well.

\begin{figure}[h!]
\centering
\includegraphics[width=0.48\textwidth]{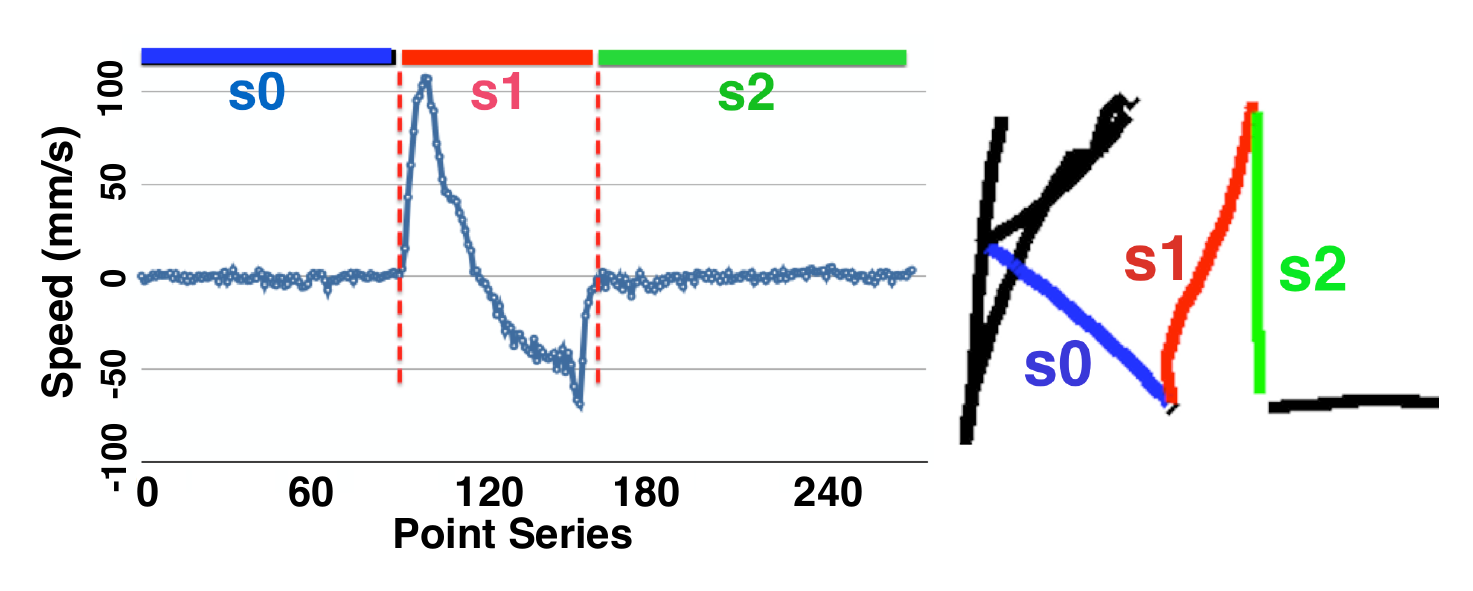}
\caption{Two connected characters "K" and "L" can be segmented by observing big jump transition in Z-direction speed diagram. One stroke in K is marked as s0, one stroke in L is marked as s2, and the bridge stroke is marked as s1. Although s1 is simply a handwritten trajectory in the air, it is recorded by the infrared sensor. However, the Z-direction speed variation is large for s1 because the writer tends to start a new stroke by lifting his/her writing tool at that period.}
\label{fig.Figure6}
\end{figure}

In order to calculate sharp angle transition from a sequence of data points, formula (1) is used to compute transition angles of every three consecutive points. An example is given to split character "Z" into three straight strokes in Fig 7. This segmentation method is often used when a writer writes a new stroke without lifting his/her tools.

\begin{equation}
\textstyle{
angle(i)=cos^{-1}  \frac{(p[i]-p[i-1])^2+(p[i]-p[i+1])^2-(p[i+1]-p[i-1])^2}{2\cdot|p[i]-p[i-1]|\cdot |p[i]-p[i+1]|}
}
\end{equation}

\begin{figure}[h!]
\centering
\includegraphics[width=0.40\textwidth]{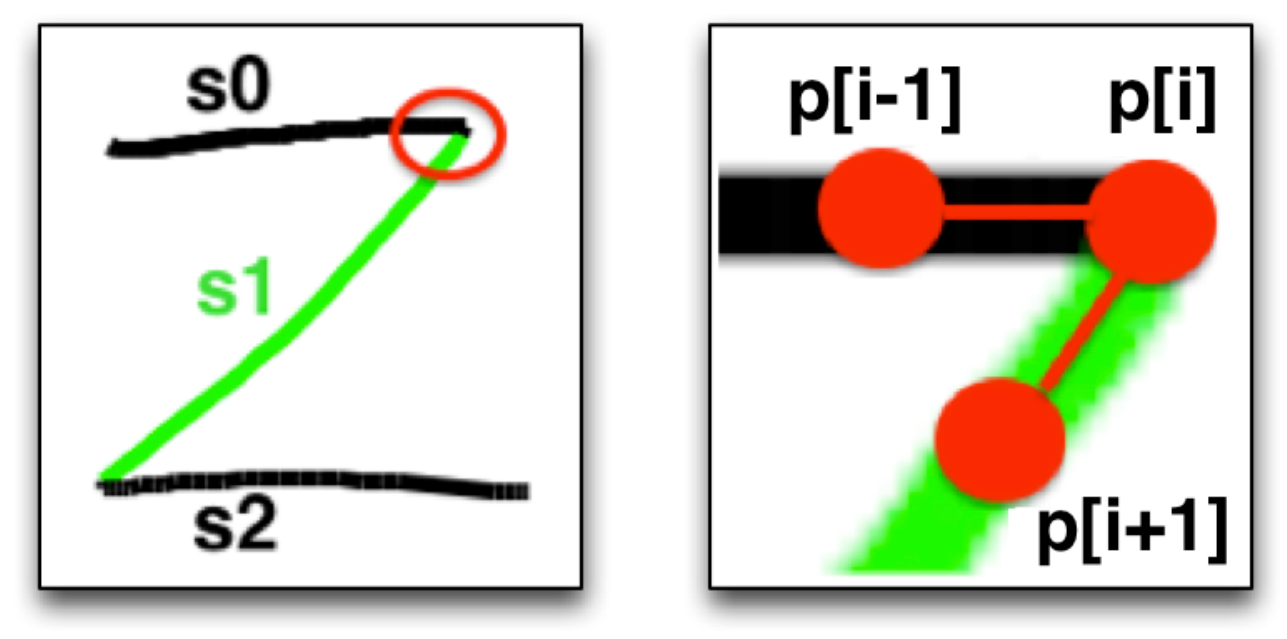}
\caption{Character "Z" can be segmented into three straight strokes based on a sharp angle detection formula. Three consecutive points can form an angle if they are not in a line. The velocity change of character "Z" from s0 to s1 is more than 120 degree which can be viewed as a sharp angle.}
\label{fig.Figure7}
\end{figure}

\subsubsection{Classification}
Classification step is targeted on identify whether a stroke is a trajectory in the air or on the board. Redundant strokes caused by off-board hand movement should be removed. After segmentation, a list of strokes are generated. However, this list contains not only on-board hand-writing, but also off-board hand movement. One intuitive solution is to check the distance of every sampled data point between the writing tool and the black/white board (commonly called depth information) when a lecturer is writing. Nevertheless, due to the limited accuracy of the infrared sensor, only a set of discontinue or missing strokes should be returned. This intuitive solution may even break the segmented strokes. To overcome this problem, strokes are utilized as the fundamental unit for binary classification. Instead of measuring depth information of every single data point, moving window standard deviation (similar to \cite{liu2015breathsens}) of the depth information in a stroke is calculated. (Fig. 8) We observe that the depth changes of on-board writing is insignificant and standard deviation calculation can remove measurement error caused by limited sensor accuracy. On the contrary, the depth changes of off-board hand moving can be considerable prominent; therefore, binary classification with training or simple threshold can effectively make reasonable judgements.

\begin{figure}[h!]
\centering
\includegraphics[width=0.48\textwidth]{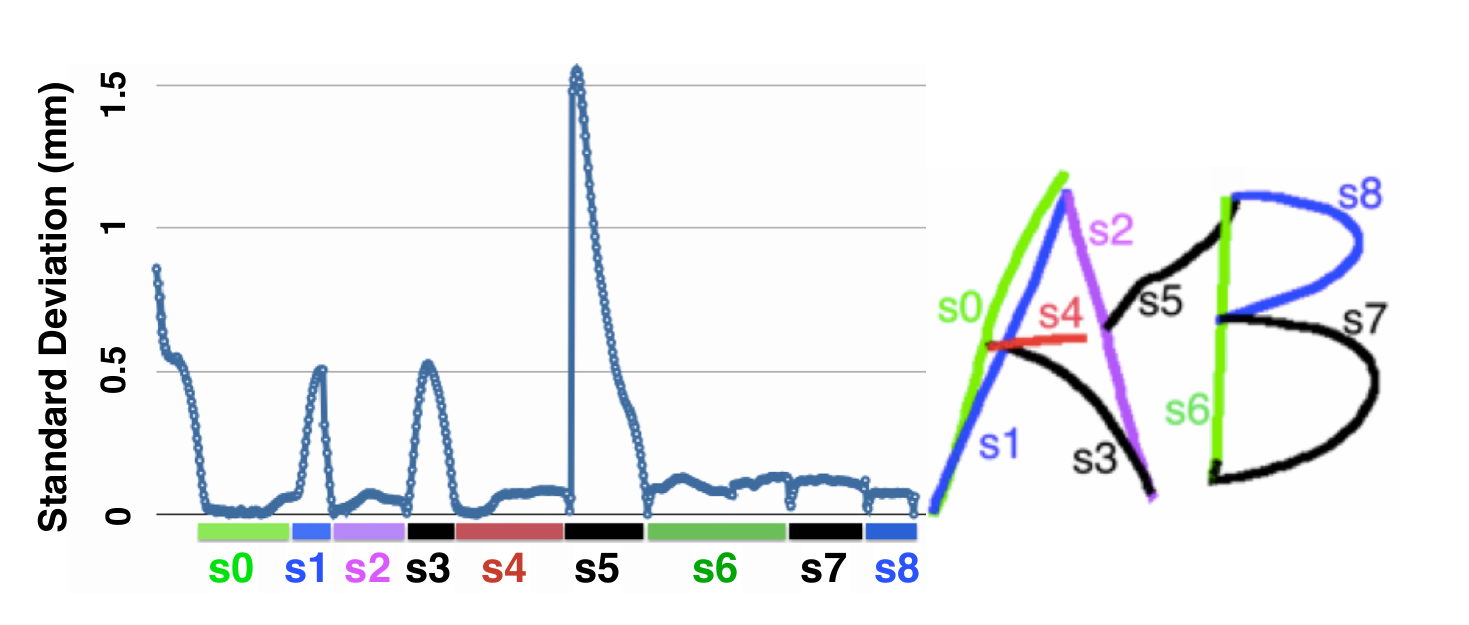}
\caption{s1, s3, and s5 are three strokes caused by off-board hand movements. In the depth standard deviation diagram, these strokes contain larger variance. On-board writing has much small depth deviation because a writing tool should be closely against to the board in board writing. From the Z direction standard deviation information, s1, s3, and s5 can be identified and removed from the segmented strokes list.}
\label{fig.Figure8}
\end{figure}

\subsubsection{Grouping}

Grouping step attempts to group a set of on-board strokes into letters if possible. This is an optional step for image-based notes generation but a required step for character recognition. A distance between every two consecutive strokes is used to test if these two strokes are belong to two characters.  A recursive method (Grouping) is used to split two neighbor characters and connect all strokes within an alphabet after splitting.\\\\\\

\begin{algorithmic}
\Function{Grouping}{$strokeList$}
    \State Pick largest distance from s[i] to s[i+1]
    \For {$j = 0 \to $i}
        \State Find rightmost point $pR \ in \ $s[j]
    \EndFor
    \For {$k = i+1 \to $n}
        \State Find leftmost point $pL \ in \ $s[k]
    \EndFor
    \If {$pR.x \leq \ $pL.x}
        \State Split strokes into two groups Left(s[0]...s[i]) and Right(s[i+1]...s[n])
        \State GROUPING(Left)
        \State GROUPING(Right)
    \Else
        \State break
    \EndIf
\EndFunction
\end{algorithmic}

\subsection{Class Note Generation}

Output from the data processing module will be images. If we apply grouping function and then recognize letters, the output will be texts. This section describes a sequence of process to generate lecture notes from generated images or texts. Some terminologies are defined below.

\begin{enumerate}
\item Notes:
Notes stand for all class notes Pages for a given class and date.
\item Pages:
Pages stand for the contents written on a column of the black/white board. Pages should be assembled by a lists of Lines and Masks with their geometric and chronical information.
\item Lines:
Every time a writer finishes a row on the board, a snapshot image (called a Line) should be created with on-board writings.
\item Masks:
Every time a writer does a modification on a created Line, a Mask will be generated to record new writing on the modified location.
\end{enumerate}

Note generation is event-driven. InfraNote system keeps track of Lines creation and Pages creation events according to English writing convention, which is writing from left to right and from top to bottom. Mask creation events are triggered when a lecturer overrides or erases written notes occasionally.

\subsubsection{Detect New Lines and New Pages Action}

\begin{figure}[htb]
\includegraphics[width=0.20\textwidth]{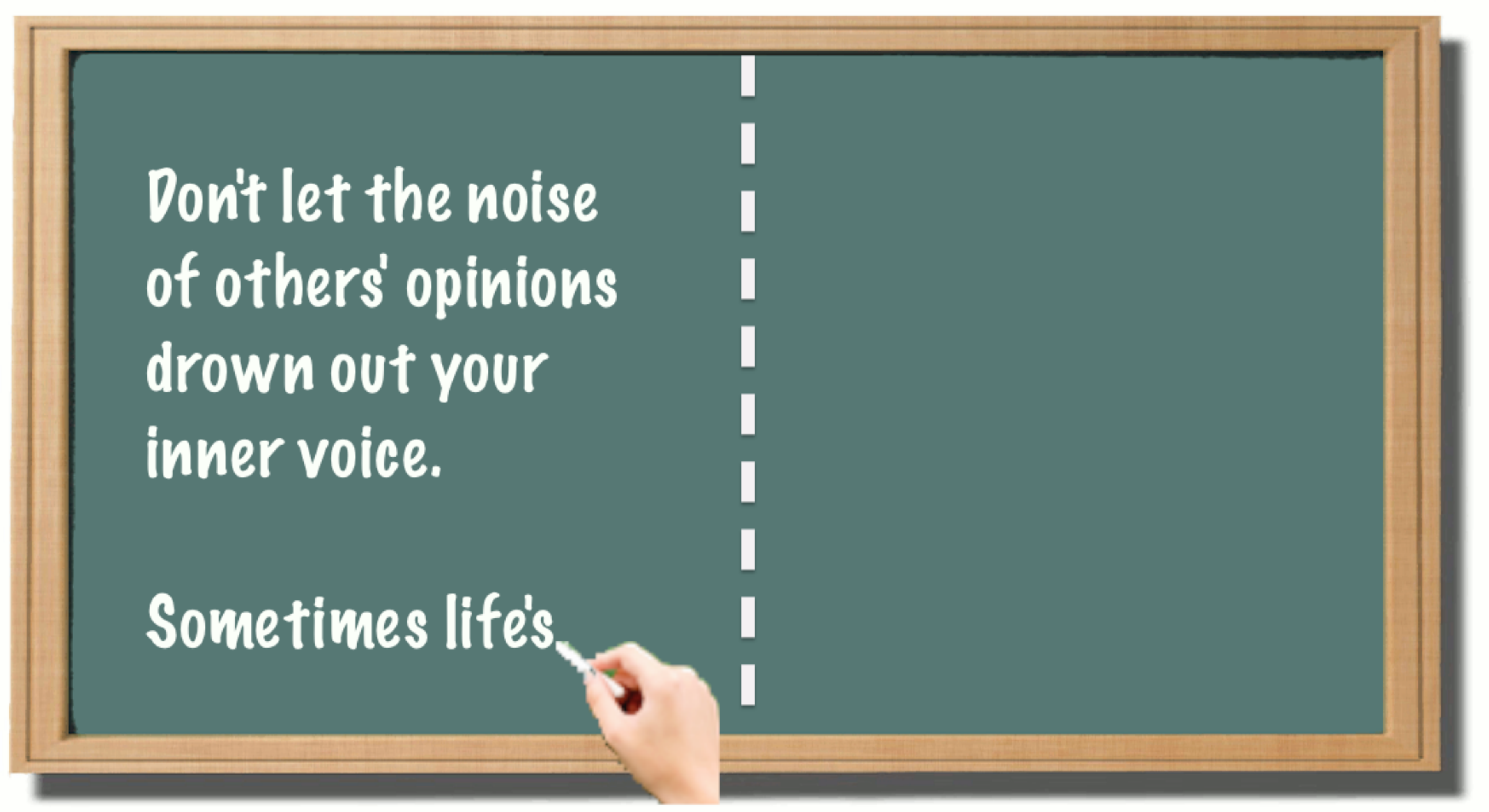}
\includegraphics[width=0.20\textwidth]{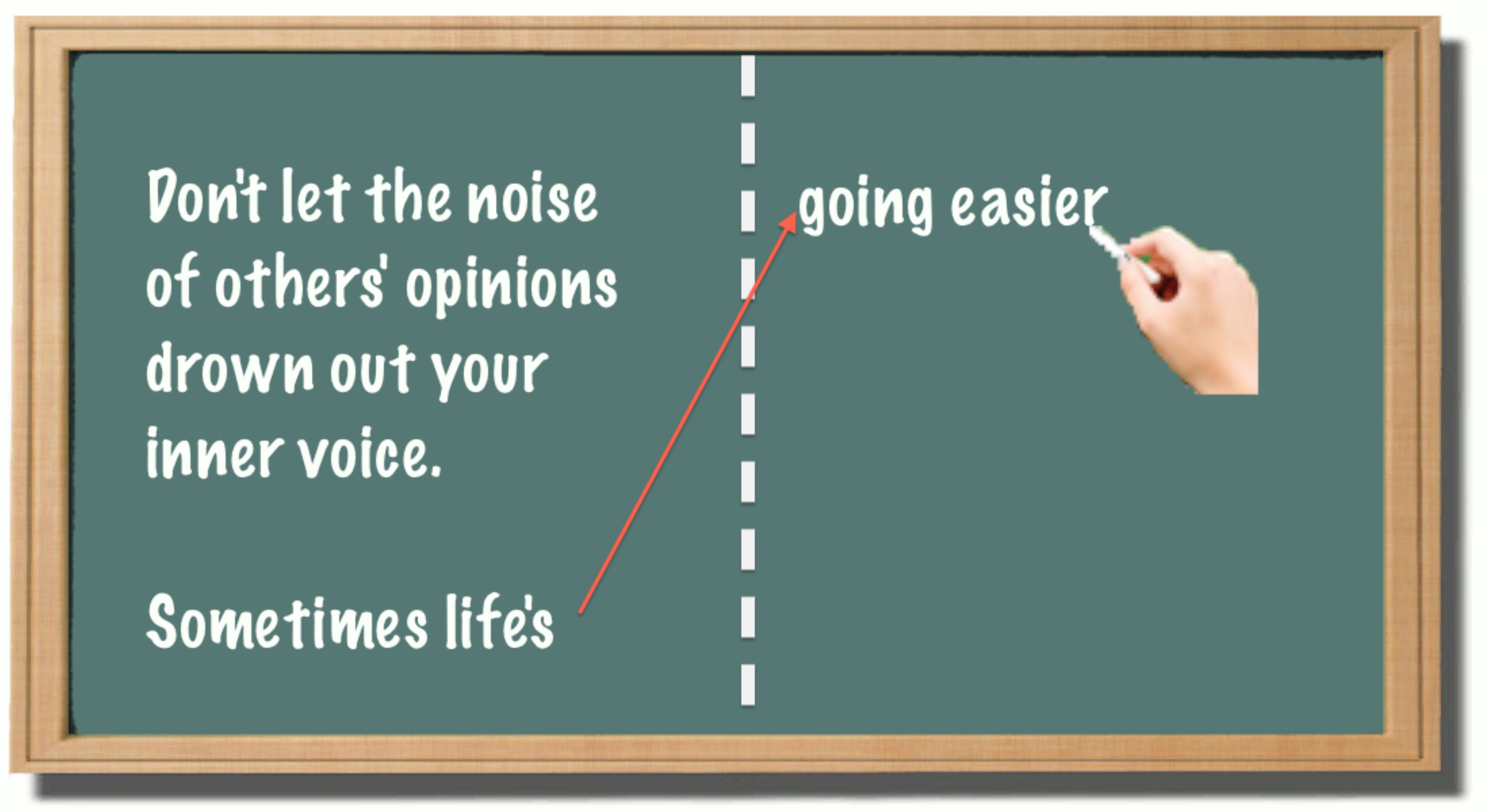}
\label{fig.Figure9}
\caption{A common notes writing convention is described: English sentence should start from left to right, and from top to bottom. When a writer finishes a sentence or reaches the border, he/she should move back to the left and start a new line below the previous line. In addition, once a writer fills a column of the board, he/she should move from bottom to the top of the board and write a new page. Both right-to-left and bottom-to-top movements are treated as special events for notes generation.}
\end{figure}

New Lines and Pages are created based on the knowledge of English writing convention: the writing order is commonly from left to right and from top to bottom. A Page should contains all contents filled in a column of the black/white board. (Fig. 9) Based on these assumptions, a new Line should be created when the writing apparently moves from the right side to the left side. Similar rule can be applied to track a new Page generation. If a writer apparently performs a movement from bottom of the board to the top level, a new Page should be generated accordingly.

\subsubsection{Detect New Masks Action}

Erasing or modifying written words and sentences are common in a lecture. Every modification should be reflected on the updated notes accordingly. LeapMotion is specially designed for tracing thin objects, such as fingers or chalks, so the eraser will be treated as two objects moving in parallel in the returned data stream. From experiments, we find that when an erasing motion happens, the LeapMotion API returns two points at a time. These points indicate the bottom of the eraser in every frame. This is a reasonable indication of note modification. In order to prevent false alarm of erasing event caused by off-board hand movement, erasing trajectories should also be passed through the proposed data processing algorithm to remove off-board movements. The on-board erasing movement can be useful to estimate the location of the Masks. If a lecturer fills new contents in the erased area, InfraNote should fill those contents into a new Mask rather than creating a new Line.

\begin{figure}[h]
\centering
\includegraphics[width=0.48\textwidth]{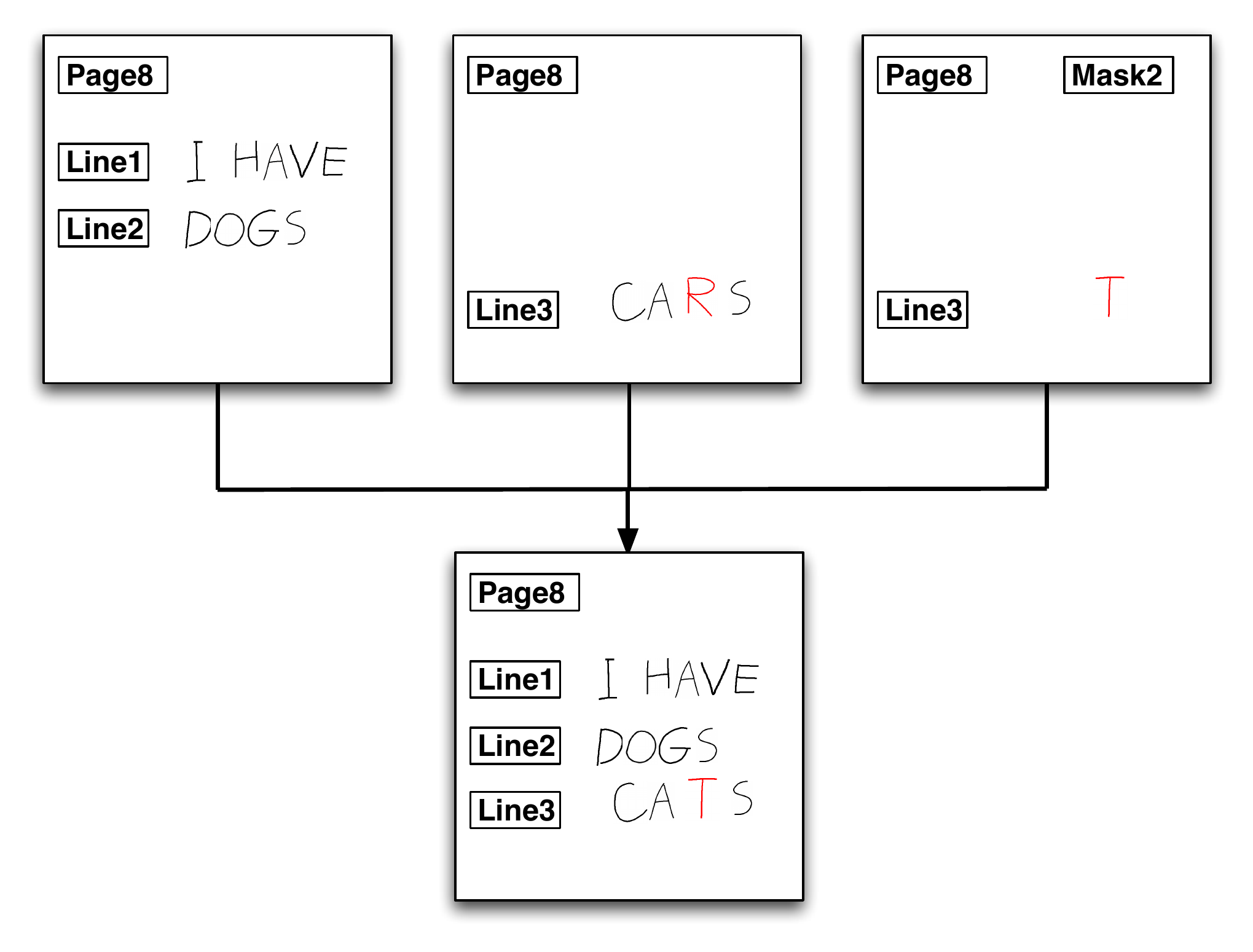}
\caption{There are two Lines (Line1 and Line2) already in Page8. Line3 is appended below according the geometric information contained in the Line. Mask2 is generated due to a detected modification event. It contains the geometric information regarding which Line it should patch to. The final version of the Page8 is a combination of Line1, Line2, Line3 with a patched content of Mask2.}
\label{fig.Figure10}
\end{figure}

\subsection{Lecture Note Assembling}

A complete lecture note is composed by several Pages. Each Page is assembled by Lines and Masks. Fig. 10 provides an example to describe how to form a page based on Lines and Masks. Each Line and Mask contain its time stamp and geometric information on the black/white board. With time stamp, each page of the lecture note can be assembled under version control. Every update will trigger a Mask creation event. Masks can be overlayed on top of the existing Lines in order to display modifications. Moreover, by applying Mask in chronical order, a Page can contain multiple versions. Default option is the most up-to-date modifications applied.

\subsection{Off-line Video Segments Indexing}

InfraNote can be integrated with a video lecture recording service. Thus, it helps students to search lecture contents from notes and video clips. Since every Page contains time stamp, once the taped video is synchronized with the InfraNote system, students can find the corresponding video clips when a lecture note is selected. This mechanism is also applicable in the reverse direction. Students can also find the corresponding lecture note Page based on video time. If character recognition is applied, students can further lookup their interested lecture notes or video clips by keywords.

\section{Notes Quality Evaluation}

Evaluating readability of the lecture notes is not a trivial task. After all, one writing style looks acceptable for somebody may not be acceptable in general. Since people are used to read typed texts either on newspaper or computers, we believe that using the criteria to identify typed text to evaluate our notes quality can be effective and reasonable. However, handwritten texts has different features than typed texts, we integrate InfraNote with online available handwritten OCR software\cite{Cite21} to evaluate the handwritten notes quality. In addition, we propose other enhanced metrics to boost OCR recognition rate by using stroke number heuristics, and primitive table methods. These heuristics and primitives are collected in the data processing stage as a side product. They are not used for note generation and assembling but used for enhance character recognition results. With sufficient accuracy in character recognition, keyword based search can be a feature of the InfraNote system. In addition to notes readability analysis, we also use these character recognition tools to analysis the robustness of InfraNote under a variety of writing styles, such as different size of texts, tilted texts, and different spaces between two consecutive characters.

\subsection{Data Stream Processing Evaluation with Character Recognition Tools}

\begin{table}[h!]
\begin{center}
\renewcommand{\thempfootnote}{\arabic{mpfootnote}}
\renewcommand{\arraystretch}{1.5}
\begin{tabular}
{|p{2.6cm}|p{1.4cm}|p{1.4cm}|p{1.4cm}|}
    \hline
{\bf }   &   {\bf OCR\footnote{Optical Character Recognition} } & {\bf OSN\footnote{OCR+Stroke Number} } & {\bf PT\footnote{Primitive Table} } \\ \hline
Raw & ~0\% & ~0\% & ~0\% \\ \hline
Segmentation & ~0\% & ~0\% &26.9\%  \\ \hline
Classification & 67.3\% & 80.8\% &34.6\%  \\ \hline
Grouping & 78.8\% & 90.4\% &92.3\%  \\ \hline
\end{tabular}
\caption{Every intermediate steps in data processing stage are analyzed by the character recognition tools. In sum, raw input without any processing cannot be recognized by any testing tools. This is because the artifacts caused by recorded off-board hand movement severely change the appearance of each alphabet. The recognition results become better after classification stage. (redundant artifacts are removed) The recognition accuracy is even improved after grouping, especially for primitive table based method, because failing to recognize individual primitives reliably will result in no-matching or incorrect matching in the primitive table.}
\end{center}
\end{table}

OCR software are commonly used for character recognition. Texts typed by computers can be identified reliably in most commercialized OCR software. Handwriting OCR is more challenging because human beings tend to have distinguishable writing styles and habits. Off-line recognizing characters captured from the black/white board is even harder because the board surface might not be clean enough. InfraNote capture notes by tracking chalks or markers movement to circumvent dirty board surface problem, but different artifacts are created because both on-board writing and off-board hand motion are loyally recorded. In fact these redundant information causes much more interference for all character recognition tools. However, in a series of experiments introduced below, we demonstrate the possibility of having reasonable recognition results with appropriate data processing and character recognition tools. Three character recognition tools are used: online available OCR software, stroke number heuristics, and primitive table.

\subsubsection{Pure OCR}

The online available handwritten OCR tools work well on typed characters and even handwritten texts on a clean paper. However, it has obvious difficulties in recognizing raw data collected by InfraNote because of the redundant artifacts caused by off-board hand movements. Once the artifacts are removed after classification the accuracy is boosted to ~70\%. (Table. 1) The accuracy can be even improved with grouping. However, the the best accuracy we can achieve is ~80\%. This is because the variance of handwritten style is huge and not the OCR library we used covers more than English words. Therefore, non-English words may be returned to lower the accuracy rate. Table. 1 also provides an example of data stream input "ABC". Pure OCR method will recognize these texts as "4BD".

\subsubsection{OCR and Stroke Number}

\begin{table}[h!]
\begin{center}
\includegraphics[width=0.30\textwidth]{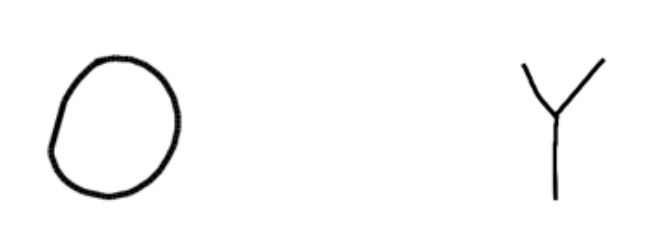}
\renewcommand{\arraystretch}{1.3}
\begin{tabular}
{|p{1.5cm}|p{1.7cm}|p{1.5cm}|p{1.7cm}|}
\hline
{\bf Candidate}   &   {\bf Probability} & {\bf Candidate} & {\bf Probability} \\ \hline
D & 0.328 & T & 0.203 \\ \hline
\textcolor{red} O & \textcolor{red}{0.319} & V & 0.201  \\ \hline
0 & 0.277 & \textcolor{red}Y & \textcolor{red}{0.156} \\ \hline
9 & 0.231 & 7 & 0.138  \\ \hline
\end{tabular}
\caption{A probability list of the pure OCR for character "O" and "Y" shows that even if the first candidates which are "D" and "T" in the lists are not the correct option, we should still be able to capture "O" and "Y" with the knowledge of number of strokes by walking through the list from the top of the list. In fact, "O" (single stroke) can be identified in the first column because "D" is written in two strokes. "T" and "V" in the second column is also written in two strokes and only the third candidate "Y" contains three strokes. Once the number of strokes heuristics match with an entry of the list, that character is returned}
\end{center}
\end{table}

\begin{table}[h!]
\begin{center}
\includegraphics[width=0.30\textwidth]{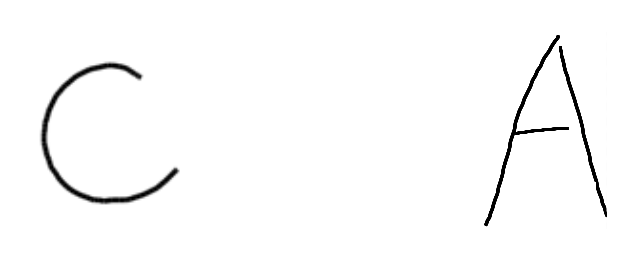}
\renewcommand{\arraystretch}{1.3}
\begin{tabular}
{|p{1.5cm}|p{1.7cm}|p{1.5cm}|p{1.7cm}|}
\hline
{\bf Candidate}   &   {\bf Probability} & {\bf Candidate} & {\bf Probability} \\ \hline
D & 0.241 & 4 & 0.190 \\ \hline
L & 0.218 & B & 0.096  \\ \hline
B & 0.196 & H & 0.095 \\ \hline
U & 0.190 & S & 0.094  \\ \hline
E & 0.174 & P & 0.082 \\ \hline
O & 0.129 & F & 0.058  \\ \hline
S & 0.111 & R & 0.052 \\ \hline
R & 0.097 & \textcolor{red}A & \textcolor{red}{0.009}  \\ \hline
\textcolor{red}C & \textcolor{red}{0.882} & L & 0.005  \\ \hline
\end{tabular}
\caption{A probability list of the pure OCR for character "C" and "A" cannot be recognized with OCR plus the number of strokes. "U" (one stroke) can be selected before "C" is met. "4" (three strokes) can be selected before "A" is met.}
\end{center}
\end{table}

The handwritten OCR software we utilized generates probability lists for each input character. The probability list changes with different writing styles and input quality. Through filtering out inappropriate candidates existing in the list, it is possible to have better recognition results with reliable heuristics. Stroke number is chosen because it can be reliably determined in the data processing stage. This heuristics should simply improve the recognition rate rather than make it worse because if a character is on the OCR software, its stroke number should be agree with the strokes we sensed from the sensor. As a result, stroke number is a reasonable candidate to be exploited to walk through the OCR probability list to search for correct characters. Table. 2 provides two examples to pick up characters "O" and "Y" from the OCR probability list with the heuristics of stroke numbers.

Table. 3 shows two examples when OCR plus stroke number heuristics fails. It is because the ranks of "C" and "A" are very low in the list. Although we may be able to filter out the first candidate "4" in the third column because it is not an alphabet, but "F" in the list is written in three strokes will be selected next.

\subsubsection{Primitive Table}

\begin{figure}[h]
\centering
\includegraphics[width=0.50\textwidth]{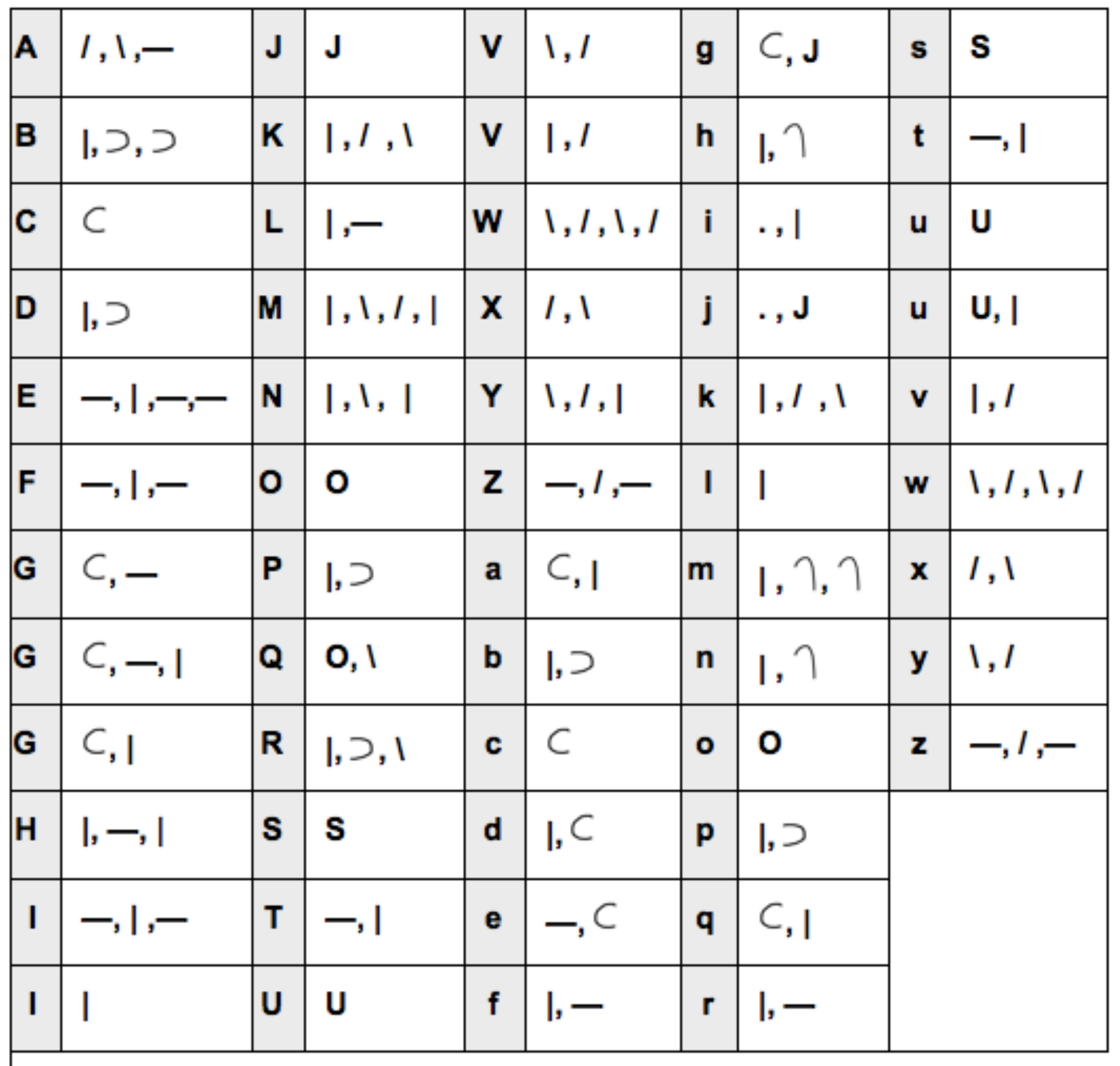}
\caption{Primitive Table: English Alphabets}
\label{fig.Figure17}
\end{figure}

A primitive table is shown in Fig. 11. It is designed to improve the recognition accuracy by enumerating all possible strokes of English alphabets by the knowledge of strokes. We decompose alphabets into several primitive type of strokes without ordering. For example, if a sequence of strokes are received as [/, \textbackslash, -], the grammar table will return "A" as the recognition result. If strokes are [/, \textbackslash, -], grammar table will also return "A".
In addition, a character can be consisted of multiple strokes combinations. Take G as an example, there are three ways to write G (Fig. 12). Whenever a new writing style is discover, the strokes combinations can be inserted into the primitive table. Nevertheless there are some of characters which are hard to categorized, such as O, S, J, and U. Hence, they are treated as basic primitives as well.

\begin{figure}[h]
\centering
\includegraphics[width=0.5\textwidth]{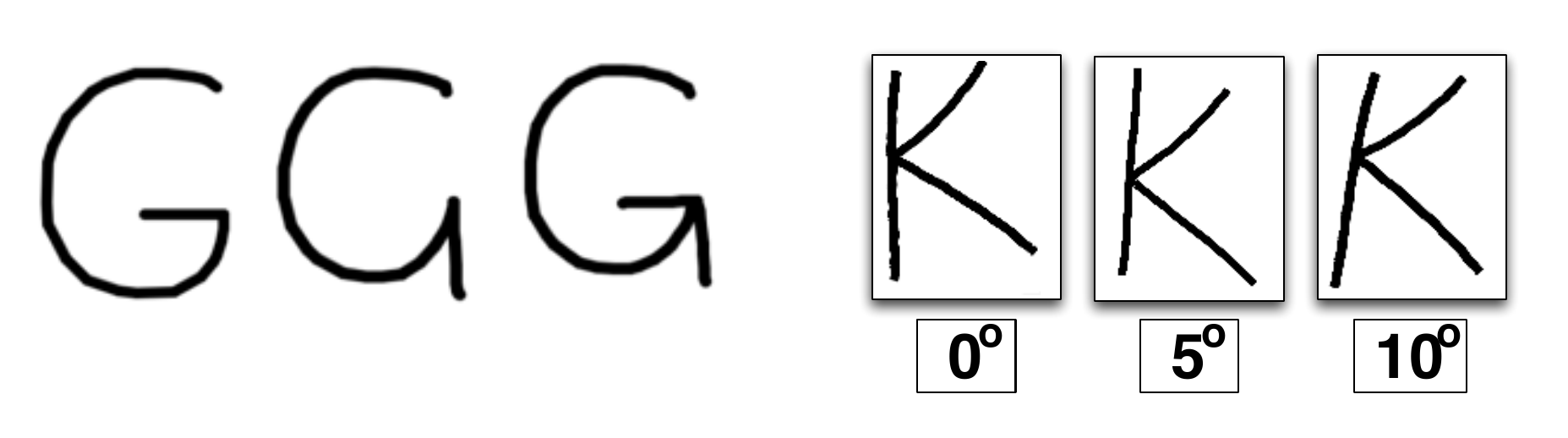}
\caption{A variant of handwritten styles of character "G". These variants should be recorded in the primitive table. Three tilted character "K" are simulated by rotating original images with 5 and 10 degree. We exploits image rotation rather than writing tilted characters on board because we want to quantify the rotated angles.}
\label{fig.Figure20}
\end{figure}

\subsection{Evaluate Tilted Characters Identification}

Writer can write characters on board with an arbitrary angle, but in general he/she may tilt within 5 to 10 degree and tile with counter-clock direction.
This phenomenon may not significant affect on the data processing step because the human behaviors in writing strokes do not depend on a tilted writing angle.
However, tilted characters tend to increase recognition error for common handwriting recognition system. To evaluate this situation, we intentionally rotate a character "K" with 5 and 10 degree counter-clock-wise (Fig. 12).  The rotated images are used as simulated inputs for pure OCR, OCR with stroke number heuristics, and primitive table methods. The accuracy drops apparently for OCR based methods as shown in Table. 4.

\begin{table}[h!]
\begin{center}
\renewcommand{\thempfootnote}{\arabic{mpfootnote}}
\renewcommand{\arraystretch}{1.5}
\begin{tabular}{|p{1.7cm}|p{1.7cm}|p{1.7cm}|p{1.7cm}|}
    \hline
{\bf }   &   {\bf OCR\footnote{Optical Character Recognition} } & {\bf OSN\footnote{OCR+Stroke Number} } & {\bf PT\footnote{Primitive Table} } \\ \hline
Original & 78.8\% & 90.4\% &92.3\% \\ \hline
5\textdegree \ Tilted & 76.9\% & 84.6\% &92.3\%  \\ \hline
10\textdegree \ Tilted & 65.4\% & 80.7\% &92.3\%  \\ \hline
\end{tabular}
\caption{In general, OCR based methods are highly affected by the tilted angles, but primitive table provides better tolerance of the uncertainty of the handwritten angles and primitive table always provides better accuracy.}
\end{center}
\end{table}

\subsection{Evaluate Identification Rate Changes Caused by the Distance between Characters}

In most of our experiments, we leave 10mm (25\% of average letter width) spacing between characters. When there are larger space between characters, results from the grouping function are better. As writer has to move further between two letters, handwritten behaviors in segmentation (slow start/end and big jump) and classification (Z-direction variance) are more clear in the waveforms. On the other hand, if there are merely 5mm spacing between characters, it will dramatically increase difficulties in grouping and severely affects the effect of the primitive table method.

\section{Discussion}

Four difficulties and design tradeoffs are brought into discussion: Sensor Range, OCR Accuracy, Modification Event Detection, and Synchronization problems

\subsection{Sensor Range}

This is our prototype using LeapMotion developer version; hence, the infrared sensor's range is very limited. One sensor cannot fully cover enough range for a full sentence. As a result, multiple LeapMotion sensors are used in the experiments with careful arrangement to avoid overlapping to their view fields. This problem should be alleviated when commercialized products or customized hardware design are available in the future.

\subsection{OCR Accuracy Limitation}

There are some pairs of characters share with the same strokes primitive combinations; for example character "D" and "P". In this case, the grammar table will return both candidates D and P and we can only use pure OCR probability list as the tie breaker. To compensate the consequence of the incorrect recognition, dictionary lookup may be necessary.

\subsection{Modification Event Detection}

Erasing action is a signal for modification. Erasing events can be detected when both sides of eraser are detected. In this case, the sensor will return two points per frame which indicate the bottom of the eraser. However, the two points per frame pattern can be simulated if two fingers or two chalks slides against the board. Although this is not common in most of lecture sensors, we believe some cheating prevention mechanism should be included in deployment.

\subsection{Synchronizing Between InfraNote System and Video Clip}

If InfraNote system and a video recording system do not launch at the same time, students cannot directly use the recorded time stamps to find the corresponding section in video. We manually keep both system synchronized in the experiments but we believe that using the recorded information of notes and video clips can achieve context based synchronization with keywords in the InfraNote and the video clips. This is one-time job for the whole lecture session; therefore, the complexity of extracting keyword from video taped notes should be acceptable.

\section{Conclusion}

This paper attempts to show a feasible prototype of an electronic lecture notes generator on the aspects of sensing and computing. Compare with other practical and theoretical solutions, our design reduces the complexity of system infrastructure while enhances user experience. Lecturers do not need to worry about adapting new sensor technology and changing their used teaching habits. All lecture notes are inconspicuously recorded because InfraNote can be seamlessly integrated with existing class room facilities: white/black board, markers and chalks. InfraNote can automatically generate and assemble class notes with our proposed data processing and software architecture design. With character recognition, created notes can integrate with lecture tapes and allow students bidirectional review course materials with bidirectional search.

\balance{}

% REFERENCES FORMAT
% References must be the same font size as other body text.
\bibliographystyle{SIGCHI-Reference-Format}
\bibliography{proceedings}

\end{document}